\documentclass[aoas,preprint]{imsart}

\RequirePackage[OT1]{fontenc}
\RequirePackage{amsthm,amsmath,bm,bbm}
\RequirePackage{amsfonts}
\RequirePackage{natbib}
\RequirePackage[colorlinks,citecolor=blue,urlcolor=blue]{hyperref}
\RequirePackage{subfigure}
\RequirePackage{graphicx} 
\RequirePackage[flushleft]{threeparttable}

\arxiv{arXiv:0000.0000}

\startlocaldefs
\numberwithin{equation}{section}
\theoremstyle{plain}

\endlocaldefs

\begin{document}

\begin{frontmatter}
\title{Development of a Common Patient Assessment Scale across the Continuum of Care: A Nested Multiple Imputation Approach\thanksref{T1}}
\runtitle{A Nested Multiple Imputation Approach}
\thankstext{T1}{Supported in part by the Changing Long-Term Care in America project funded by the National Institute on Aging (P01AG027296).}

\begin{aug}
  \author{\fnms{Chenyang}  \snm{Gu}\corref{}\thanksref{m1}\ead[label=e1]{gu@hcp.med.harvard.edu}}
  \and
  \author{\fnms{Roee}  \snm{Gutman}%
  \thanksref{m2}\ead[label=e3]{roee$\_$gutman@brown.edu}}%


  \runauthor{C. Gu and R. Gutman}

  \affiliation{Harvard University\thanksmark{m1} and Brown University\thanksmark{m2}}

  \address{Chenyang Gu\\Department of Health Care Policy\\
  Harvard Medical School\\
  Boston, Massachusetts 02115\\
  USA\\ 
          \printead{e1}}

  \address{Roee Gutman\\Department of Biostatistics\\
  Brown University School of Public Health\\
Providence, Rhode Island 02912\\
USA\\
\printead{e3}}

\end{aug}

\begin{abstract}
Evaluating and tracking patients' functional status through the post-acute care continuum requires a common instrument. However, different post-acute service providers such as nursing homes, inpatient rehabilitation facilities and home health agencies rely on different instruments to evaluate patients' functional status. These instruments assess similar functional status domains, but they comprise different activities, rating scales and scoring instructions. These differences hinder the comparison of patients' assessments across health care settings. We propose a two-step procedure that combines nested multiple imputation with the multivariate ordinal probit (MVOP) model to obtain a common patient assessment scale across the post-acute care continuum. Our procedure imputes the unmeasured assessments at multiple assessment dates and enables evaluation and comparison of the rates of functional improvement experienced by patients treated in different health care settings using a common measure. To generate multiple imputations of the unmeasured assessments using the MVOP model, a likelihood-based approach that combines the EM algorithm and the bootstrap method as well as a fully Bayesian approach using the data augmentation algorithm are developed. Using a dataset on patients who suffered a stroke, we simulate missing assessments and compare the MVOP model to existing methods for imputing incomplete multivariate ordinal variables. We show that, for all of the estimands considered, and in most of the experimental conditions that were examined, the MVOP model appears to be superior. The proposed procedure is then applied to patients who suffered a stroke and were released from rehabilitation facilities either to skilled nursing facilities or to their homes.\\
\end{abstract}



\end{frontmatter}

\section{Introduction}

\subsection{Overview}

To track and evaluate patients through the post-acute care continuum a common standardized evaluation tool is needed. Current evaluation tools have been largely developed within each type of health care provider, and cannot be easily compared. In inpatient rehabilitation facilities (IRFs), patients' functional status is evaluated by the Functional Independence Measure (FIM). After being discharged from IRFs, the functional status of patients who stay in skilled nursing facilities (SNFs) is collected using the Minimum Data Set (MDS), while the Outcome and Assessment Information Set (OASIS) is collected for patients who receive home health care provided by home health agencies. All of these assessments examine similar functional capabilities (e.g., eating, grooming, dressing, etc.), but the specific instruments, rating scales and instructions for scoring the different activities vary between these post-acute settings. Thus, it is difficult to evaluate and compare the rates of functional improvement experienced by patients treated in the different health care settings.

The Continuity Assessment Record and Evaluation item set is a standardized evaluation tool that was developed for use at acute hospital discharge and at post-acute care admission and discharge \citep{gage1development}. This tool is intended to be a common evaluation tool for evaluating patients across the continuum of post-acute care, and considerable resources were invested in its development. However, implementing new instruments in all post-acute care settings may result in additional investments in administration and training as well as in changes to the reimbursement system \citep{li2017linking}. Moreover, adopting new instruments would require translating past functional status scores so that comparison to the new scores is possible.

Equating setting-specific instruments so that functional status scores from one instrument could be used interchangeably with ones from another instrument is a possible approach to obtain a common evaluation tool \citep{kolen2004test, dorans2007linking, von2010statistical}. Linking and equating scores across different standardized assessments has been a major focus in the field of educational testing for the past 90 years \citep[see][chapter 2, for details]{dorans2007linking}. Score equating methods have been recently used in health outcomes research. The conversion table method \citep{velozo2007translating} was used to equate FIM assessments with MDS assessments. Conversion table equates the sum of individual item scores, also referred to as the total score, by matching on latent functional scores that are estimated from two different instruments using Item Response Theory models. This method was also used to equate scores from two physical functioning scales \citep{ten2013development}, as well as two depression scales \citep{fischer2011compare}. Because conversion table ignores the variability from the estimation and the imputation processes, it may result in statistically invalid estimates when further analysis is performed using the imputed scores \citep{gu2016irt}. Furthermore, a data set that comprises contemporaneous MDS and OASIS assessments is required in order to equate MDS and OASIS instruments. However, MDS and OASIS assessments are never jointly observed.

We propose a nested multiple imputation procedure \citep{shen2000nested, harel2003strategies, rubin2003nested} to impute unmeasured assessments across the continuum of care. This procedure enables evaluation and comparison of the rates of functional improvement across different health care settings, and it consists of an Equating step and a Translating step. In the Equating step, we impute the unmeasured assessments in MDS or OASIS that are close to the FIM assessment date to obtain a synthetic data set with simultaneous MDS and OASIS assessments. In the Translating step, we rely on the synthetic data set from the first step to estimate the relationship between MDS and OASIS that will be used to impute multiple unmeasured assessments in MDS or OASIS at later assessment dates. This two-step procedure accounts for the uncertainty in both steps, and provides flexibility for researchers to choose different models in the second step without the need to re-equate the instruments. 

The Equating step imputes the missing instruments that consist of multiple ordinal items. The logistic and probit link functions are commonly used to model single ordinal variable. These link functions give similar model fit and predictive performance. Bayesian inference for these models relies on sampling from complex posterior distributions. However, by introducing auxiliary variables, sampling from these posterior distributions can become more efficient. \citet{albert1993bayesian} described this technique for the probit link function and more recently \citet{holmes2006bayesian} and \citet{polson2013bayesian} proposed two possible approaches for the logistic link function. The multivariate ordinal probit (MVOP) model was proposed as an extension of the probit model \citep{albert1993bayesian} and the multivariate probit model \citep{ashford1970multi} to multivariate ordinal responses. Similar extensions for logistic link function with multiple ordinal outcomes is an area of future research. Using the MVOP model, we can capture the complex dependence structure and the ordinal nature of the different functional assessment instruments as well as adjust for observed patients' covariates. 

To generate multiple imputations of the unmeasured functional assessments using the MVOP model, we develop two computational approaches. The first approach combines the EM algorithm \citep{dempster1977maximum} for obtaining the maximum likelihood estimates of the parameters in the MVOP model and the bootstrap method to multiply impute the missing values \citep{little2002statistical}. The second approach relies on the data augmentation (DA) algorithm \citep{tanner1987calculation} to draw the unknown parameters of the MVOP model and the missing values from their joint posterior distribution. We compared the MVOP model to existing methods for imputing incomplete multivariate ordinal variables with respect to the biases, the sampling variances, and the RMSEs of their point estimates, as well as the widths and coverage rates of their interval estimates. For all of the estimands considered and in most of the experimental conditions that were examined, the MVOP model appears to be superior. In the Translating step, different models can be used to estimate the relationship between MDS and OASIS assessments. We illustrate this flexibility either by imputing the missing individual items using the MVOP model or by imputing the missing total scores using a linear regression model.

The remainder of this section describes the analytical data set, introduces the basic framework and reviews related work. Section 2 describes the MVOP models and their estimation methods. Section 3 presents the nested multiple imputation procedure. Section 4 compares the MVOP models to existing methods using a simulation study. Section 5 describes the empirical data analysis. Conclusions and discussions are provided in Section 6. 

\subsection{Motivating Example}
The analytical data set includes 72,575 patients who suffered a stroke and were discharged from IRFs between 2011 and 2014. Of these patients, 38,629 were released to SNFs, where the MDS assessments were collected for them. The other 33,946 patients were discharged home, where the OASIS assessments were used to measure their functional status. Patient assessments were collected on admission and at various time points during their post-acute stays. The median number of assessments for each patient was 5 for patients in SNFs and the range was 0 to 91. The median number of assessments for patients receiving home health care was 3 and the range was 0 to 46. Two assessments for each patient were included in our analyses. One assessment was collected at admission within 30 days from the IRF's discharge date. The other was recorded approximately 30 days after the first assessment. The primary research objective is to examine and compare the rates of functional improvement experienced by patients treated in the different health care settings after being discharged from IRFs. To describe the functional change for patients who were released to either SNFs or home, we will use the MDS scale. 

\subsection{Basic Framework}
We consider equating setting-specific patient assessments as a missing data problem. We assume that all patients have complete FIM assessments and complete demographic characteristics. Let $\mathbf{M} = \{M_{i}\}$, $i = 1,\ldots, N$, where $M_{i}$ is an indicator that is equal to 1 if patient $i$ was discharged home and 0 otherwise. Let $\mathbf{Y}^{\text{fim}}$, $\mathbf{Y}^{\text{mds}} = (\mathbf{Y}^{\text{mds}}_{A}, \mathbf{Y}^{\text{mds}}_{B})$, and $\mathbf{Y}^{\text{oas}} = (\mathbf{Y}^{\text{oas}}_{A},\mathbf{Y}^{\text{oas}}_{B})$ denote matrices of item responses in FIM, MDS, and OASIS, respectively, with rows referring to subjects and columns referring to variables, and where $\mathbf{Y}^{\text{mds}}_{A} = (\mathbf{Y}^{\text{mds}}_{A,\text{obs}},\mathbf{Y}^{\text{mds}}_{A,\text{mis}})$, $\mathbf{Y}^{\text{mds}}_{B} = (\mathbf{Y}^{\text{mds}}_{B,\text{obs}},\mathbf{Y}^{\text{mds}}_{B,\text{mis}})$, $\mathbf{Y}^{\text{oas}}_{A} = (\mathbf{Y}^{\text{oas}}_{A,\text{mis}},\mathbf{Y}^{\text{oas}}_{A,\text{obs}})$, and $\mathbf{Y}^{\text{oas}}_{B} = (\mathbf{Y}^{\text{oas}}_{B,\text{mis}},\mathbf{Y}^{\text{oas}}_{B,\text{obs}})$. The subscripts A and B denote the assessments on admission and at the later date, respectively. The subscripts \textit{obs} and \textit{mis} denote the observed and missing assessments, respectively. In addition, let $\mathbf{X}$ denote a set of fully observed covariates. 

The joint posterior distribution of the missing data and the parameters can be written as 
\small{
\begin{equation}
\begin{split}
&f(\mathbf{Y}^{\text{mds}}_{A,\text{mis}}, \mathbf{Y}^{\text{mds}}_{B,\text{mis}}, \mathbf{Y}^{\text{oas}}_{A,\text{mis}}, \mathbf{Y}^{\text{oas}}_{B,\text{mis}}, \bm{\psi}_A, \bm{\psi}_B | \mathbf{Y}^{\text{mds}}_{A,\text{obs}}, \mathbf{Y}^{\text{mds}}_{B,\text{obs}}, \mathbf{Y}^{\text{oas}}_{A,\text{obs}}, \mathbf{Y}^{\text{oas}}_{B,\text{obs}}, \mathbf{Y}^{\text{fim}}, \mathbf{X}, \mathbf{M}) \\
&= f(\mathbf{Y}^{\text{mds}}_{A,\text{mis}}, \mathbf{Y}^{\text{oas}}_{A,\text{mis}}, \bm{\psi}_A | \mathbf{Y}^{\text{mds}}_{A,\text{obs}}, \mathbf{Y}^{\text{oas}}_{A,\text{obs}}, \mathbf{Y}^{\text{fim}}, \mathbf{X}, \mathbf{M})\\
& \times f(\mathbf{Y}^{\text{mds}}_{B,\text{mis}}, \mathbf{Y}^{\text{oas}}_{B,\text{mis}}, \bm{\psi}_B| \mathbf{Y}^{\text{mds}}_{A,\text{mis}}, \mathbf{Y}^{\text{oas}}_{A,\text{mis}}, \mathbf{Y}^{\text{mds}}_{A,\text{obs}}, \mathbf{Y}^{\text{mds}}_{B,\text{obs}}, \mathbf{Y}^{\text{oas}}_{A,\text{obs}}, \mathbf{Y}^{\text{oas}}_{B,\text{obs}}, \mathbf{Y}^{\text{fim}}, \mathbf{X}, \mathbf{M}),
\end{split}
\label{equation1}
\end{equation}}where $\bm{\psi}_A$ and $\bm{\psi}_B$ index the imputation models in the Equating and Translating steps, respectively. The Equating step is performed once, and the Translating step can be performed multiple times. To reduce the computational complexity and to provide flexibility to researchers we assumed in Equation (\ref{equation1}) that $\bm{\psi}_A$ and $\bm{\psi}_B$ are conditionally independent. The data setting that consists of patients' covariates, FIM assessments, and first MDS or OASIS assessments resembles the statistical matching setup \citep{d2006statistical}. In this setup, the joint distribution of $\{ \mathbf{Y}^{\text{fim}}, \mathbf{Y}_{A}^{\text{mds}}, \mathbf{Y}_{A}^{\text{oas}}, \mathbf{X} \}$ is not identifiable based on observed data, because MDS and OASIS are never jointly observed. 

This setup also arises in the test equating literature when using common-item nonequivalent groups design \citep{kolen2004test, dorans2007linking}. This design assumes that different groups of examinees are assessed using two different test forms that share a common item set. When used for equating, the common-item set should be representative of the total test forms in content and statistical characteristics \citep{kolen2004test, dorans2007linking}. This is commonly attained by ensuring that the items are exactly the same in both forms and are at the same location in the form. Here, $\mathbf{Y}^{\text{fim}}$ is similar for all patients, and it is administered prior to and within a short time frame from the initial MDS and OASIS assessments. In addition, $\mathbf{Y}^{\text{fim}}$ includes similar content to the MDS and OASIS assessments, because it attempts to approximate the same underlying functional status. Based on these observations, a natural starting point is to apply  the conditional independence assumption, $f(\mathbf{Y}_{A}^{\text{mds}},\mathbf{Y}_{A}^{\text{oas}}|\mathbf{Y}^{\text{fim}}, \mathbf{X}, \mathbf{M}, \bm{\psi}_A) = f(\mathbf{Y}_{A}^{\text{mds}}|\mathbf{Y}^{\text{fim}}, \mathbf{X}, \mathbf{M}, \bm{\psi}_A) f(\mathbf{Y}_{A}^{\text{oas}}|\mathbf{Y}^{\text{fim}}, \mathbf{X}, \mathbf{M}, \bm{\psi}_A)$ \citep{d2006statistical}. This assumption is often implicitly made in test equating applications using only $\mathbf{Y}^{\text{fim}}$. Here, we include other patient characteristics as well. 

We further assumed that the unmeasured assessments are missing at random (MAR) \citep{little2002statistical}, because a major determinant of patients' discharge destination from a rehabilitation facility is their functional status, which is measured using the validated FIM instrument. Under the conditional independence and the MAR assumptions, we can impute $\mathbf{Y}^{\text{mds}}_{A,\text{mis}}$ using the posterior distribution $f(\mathbf{Y}_{A,\text{mis}}^{\text{mds}}|\mathbf{Y}^{\text{fim}}, \mathbf{Y}_{A,\text{obs}}^{\text{mds}}, \mathbf{X}, \bm{\psi}_A)$ in the Equating step. These two assumptions cannot be inferred from the data and may not always be plausible. To examine the plausibility of these assumptions, we conducted a sensitivity analysis to investigate whether our results changed in a substantial way when these assumptions are violated \citep{rubin1986statistical, heitjan1994assessing}. 

The Equating step generates complete synthetic data sets that comprise MDS and OASIS assessments simultaneously for patients who were discharged home. Assuming MAR and that the relationship between contemporary imputed and observed instruments does not change across the continuum of care, we can simplify the third line of Equation (\ref{equation1}):
\begin{equation*}
\begin{split}
& f(\mathbf{Y}^{\text{mds}}_{B,\text{mis}}, \mathbf{Y}^{\text{oas}}_{B,\text{mis}}, \bm{\psi}_B| \mathbf{Y}^{\text{mds}}_{A,\text{mis}}, \mathbf{Y}^{\text{oas}}_{A,\text{mis}}, \mathbf{Y}^{\text{mds}}_{A,\text{obs}}, \mathbf{Y}^{\text{mds}}_{B,\text{obs}}, \mathbf{Y}^{\text{oas}}_{A,\text{obs}}, \mathbf{Y}^{\text{oas}}_{B,\text{obs}}, \mathbf{Y}^{\text{fim}}, \mathbf{X}, \mathbf{M})\\
& = f(\mathbf{Y}^{\text{mds}}_{B,\text{mis}}, \bm{\psi}_B^{\text{mds}}| \mathbf{Y}^{\text{mds}}_{A,\text{mis}}, \mathbf{Y}^{\text{oas}}_{A,\text{obs}}, \mathbf{Y}^{\text{oas}}_{B,\text{obs}}) \times 
f(\mathbf{Y}^{\text{oas}}_{B,\text{mis}}, \bm{\psi}_B^{\text{oas}}| \mathbf{Y}^{\text{oas}}_{A,\text{mis}}, \mathbf{Y}^{\text{mds}}_{A,\text{obs}}, \mathbf{Y}^{\text{mds}}_{B,\text{obs}}),
\end{split}
\end{equation*}
where $\bm{\psi}_B = (\bm{\psi}_B^{\text{mds}}, \bm{\psi}_B^{\text{oas}})$, and in the Translating step we impute $\mathbf{Y}^{\text{mds}}_{B,\text{mis}}$ using $f(\mathbf{Y}_{B,\text{mis}}^{\text{mds}} |\mathbf{Y}_{A,\text{imp}}^{\text{mds}}, \mathbf{Y}_{A,\text{obs}}^{\text{oas}}, \mathbf{Y}_{B,\text{obs}}^{\text{oas}}, \bm{\psi}_B^{\text{mds}})$, where $\mathbf{Y}_{A,\text{imp}}^{\text{mds}}$ denotes the imputed MDS assessments at admission.

\subsection{Related Work}
The Equating and the Translating steps require methods that impute multivariate ordinal variables. Multivariate imputation methods can be classified into two types of methods: fully conditional specification and joint modeling. Fully conditional specification \citep{van2007multiple} involves a series of univariate conditional models that imputes missing values sequentially with current model estimates. In practice, users only include main effects in these models, because it is challenging to identify and include higher-order interactions and nonlinear terms at each of the conditional models \citep{vermunt2008multiple}. With multiple ordinal variables, the default implementation of fully conditional specification relies on the ordered logit model. \citet{gu2016irt} noted that this implementation fails to capture the full correlation structure of the imputed items when the proportional odds assumption is violated. Recently, a multi-level model based on the probit link function was proposed as a possible imputation model for missing ordinal variable \citep{enders2017fully}.


The joint modeling approach \citep{schaferanalysis} specifies a joint probability model for all the data. Imputation of missing values is performed from the implied distribution of the missing variables conditional on the observed data. \citet{yucel2011gaussian} proposed a method that is based on multivariate normal model to impute ordinal variables and supplemented it with a rounding technique that preserves the observed marginal distribution of the ordinal variables. When there is a large proportion of missing values, propagation of errors in the underlying modeling approximation can compound and result in invalid statistical inferences \citep{yucel2011gaussian, gu2016irt}. 

Imputation by Propensity score matching (IPSM) can be embedded in a joint modeling approach to define cells within which hot-deck imputations can be drawn \citep{andridge2010review}. The propensity score is defined as the probability of a unit to have missing values. IPSM imputes missing values with observed values from units with similar estimated propensity scores. IPSM is a generally valid statistical method, but its performance is sensitive to the specification of the propensity score model \citep{gu2016irt}.

Latent variable matching (LVM) \citep{gu2016irt} is a recently proposed procedure that combines IRT models with multiple imputation \citep{rubin1987multiple} to impute unmeasured assessments. LVM is also a hot deck imputation method, which matches units using the underlying functional status estimated from IRT models. In its original form, LVM ignores patient covariates, which may violate the MAR assumption \citep{rubin1996multiple}. LVM can be extended to account for a set of discrete and continuous covariates by applying it within subgroups of the covariates; however, this approach may become computationally intensive when the number of possible covariate values is large.

Among these methods, IPSM and LVM are the strongest candidate methods in terms of validity and efficiency for imputing the missing assessments in our datasets \citep{gu2016irt}. Thus, we only compared these two methods with the newly proposed procedure in Section 4.

\section{The Multivariate Ordinal Probit Model}
Let $\mathbf{Y} = (\mathbf{Y}_{\text{obs}}, \mathbf{Y}_{\text{mis}}) = \{y_{ij}, i = 1,\ldots,N, j = 1,\ldots,J\}$ denote a generic matrix of item responses, where $\mathbf{Y}_{\text{mis}}$ and $\mathbf{Y}_{\text{obs}}$ are the matrices of missing and observed item responses, respectively, $y_{ij} \in \{1,\ldots,c_j\}$ is the response of patient $i$ to item $j$, and $c_j$ is the number of response levels of item $j$. For example, in the Equating step, $\mathbf{Y}_{\text{mis}}$ corresponds to the unmeasured assessments in MDS, $\mathbf{Y}_{A,\text{mis}}^{\text{mds}}$, and $\mathbf{Y}_{\text{obs}}$ corresponds to the observed assessments in FIM and MDS, $\mathbf{Y}^{\text{fim}}$ and $\mathbf{Y}_{A,\text{obs}}^{\text{mds}}$. The MVOP model introduces a matrix of latent response variables $\mathbf{Z} = \{z_{ij}, i = 1,\ldots,N, j = 1,\ldots,J \}$ such that $y_{ij} = g_j(z_{ij}) = l$, if $\gamma_{j,l-1} < z_{ij} \leq \gamma_{j,l}$, where $-\infty = \gamma_{j,0} < \gamma_{j,1} < \cdots < \gamma_{j,c_j-1} < \gamma_{j,c_j} = +\infty$ are unknown threshold parameters. The MVOP model assumes that $\mathbf{z}_i = (z_{i1}, \ldots, z_{iJ})^{\top} \sim \mathcal{N}(\bm{\beta} \mathbf{x}_i, \bm{\Sigma}), i = 1,\ldots,N$, where $\mathbf{x}_i$ is a $P \times 1$ vector of covariates for patient $i$, $\bm{\beta}$ is a $J \times P$ matrix of unknown regression coefficients and $\bm{\Sigma}$ is an unknown covariance matrix. Statistical inferences for $\bm{\psi} = (\bm{\gamma}, \bm{\beta}, \bm{\Sigma})$ are based on the likelihood 
\begin{equation}
\begin{split}
L(\bm{\psi} | \mathbf{Y}_{\text{obs}}, \mathbf{X}) &=  c \prod_{i=1}^N \int  f(\mathbf{y}_i | \mathbf{x}_i, \bm{\psi}) d\mathbf{y}_{\text{mis},i}\\
&= c \prod_{i=1}^N \int \int_{\Gamma_{iJ}} \cdots \int_{\Gamma_{i1}} \mathcal{N}_{J}(\mathbf{z}_i; \bm{\beta}\mathbf{x}_i, \bm{\Sigma}) d\mathbf{z}_i d\mathbf{y}_{\text{mis},i},
\end{split}
\label{equation2}
\end{equation}
 where $\Gamma_{ij}$ is the interval $(\gamma_{i,l-1}, \gamma_{i,l}]$ if $y_{ij} = g_j(z_{ij}) = l$, and $c$ is a normalizing constant.

The vector parameter $\bm{\psi}$ is not identifiable because the likelihood (\ref{equation2}) is invariant to location and scale transformations on $\mathbf{Z}$. Threshold constraints and correlation constraints are two types of identification constraints that are commonly made with the MVOP model. The threshold constraints fix two threshold parameters for each outcome. For example, one could set $\gamma_{j,1} = 0$ and either $\gamma_{j,2} = 1$ or $\gamma_{j,c_j-1} = 1$ $\forall j$. Applying these constraints allows to sample the covariance matrix $\bm{\Sigma}$  from a known probability distribution \citep{chen2000unified, jeliazkov2008fitting}. The correlation constraints either fix $\gamma_{j,1} = 0$ or $\beta_{j,1} = 0$ $\forall j$, and restrict $\bm{\Sigma}$ to be a correlation matrix $\mathbf{R}$ \citep{lawrence2008bayesian, zhang2016multiple}. We refer to the MVOP model under the threshold constraints and correlation constraints as the MVOPT model and MVOPC model, respectively.

MVOP models have been analyzed using likelihood-based methods, including a direct likelihood approach that involves the evaluation of integrals using Gaussian-Hermite quadrature \citep{li2008likelihood}, an approximate EM algorithm \citep{guo2015graphical}, a pseudo-likelihood approach \citep{varin2009mixed}, and Bayesian approaches \citep{chen2000unified, lawrence2008bayesian, zhang2016multiple}. Of these methods, only \citet{zhang2016multiple} extended the MVOP model to handle incomplete correlated ordinal responses. Here, when some of the outcomes are missing, we propose Monte Carlo EM (MCEM) algorithms \citep{wei1990monte} for maximum likelihood estimation and DA algorithms for Bayesian inference under the MVOP models to produce imputed values.

\subsection{MCEM Algorithm}
We first consider the MVOPT model, and fix $\gamma_{j,1} = 0$ and $\gamma_{j,c_j-1} = 1$ $\forall j$. The complete-data likelihood is
\begin{equation*}
\begin{split}
L_{\text{com}}(\bm{\psi} | \mathbf{Y}, \mathbf{X}, \mathbf{Z})  &\propto  \left| \bm{\Sigma} \right|^{-\frac{N}{2}} 
 \exp\Big{\{} -\frac{1}{2} \mathrm{tr}(\bm{\Sigma}^{-1} \sum_{i=1}^N (\mathbf{z}_i - \bm{\beta}\mathbf{x}_i)(\mathbf{z}_i - \bm{\beta}\mathbf{x}_i)^{\top}) \Big{\}}  \\
 & \times \prod_{i=1}^N\prod_{j=1}^J \bm{1}\{ z_{ij} \in \Gamma_{ij} \}.
 \end{split}
\end{equation*}

The E-step of the EM algorithm, given the current value of the parameter, $\bm{\psi}$, involves evaluating the expectation
\begin{equation*}
\begin{split}
Q(\bm{\psi}^{*} | \bm{\psi}) &= E\{ \log L(\bm{\psi}^{*} | \mathbf{Y}, \mathbf{X}, \mathbf{Z}) | \mathbf{Y}_{\text{obs}}, \mathbf{X}, \bm{\psi}\} \\
&= \int \log L(\bm{\psi}^{*} | \mathbf{Y}, \mathbf{X}, \mathbf{Z}) f(\mathbf{Z} | \mathbf{Y}_{\text{obs}}, \mathbf{X}, \bm{\psi}) d\mathbf{Z},
\end{split}
\end{equation*}
which consists of multiple integrations with respect to a truncated multivariate normal distribution of $\mathbf{Z}$. $Q(\bm{\psi}^{*} | \bm{\psi})$ cannot be calculated analytically, but Monte Carlo methods can be used to approximate it. We extend the slice sampler algorithm proposed by  \citet{damien2001sampling} for bivariate normal distribution to sample from the truncated multivariate normal distribution. The algorithm introduces a latent variable so that the slice sampler runs on a sequence of conditional distributions which can all be sampled directly using uniform distributions. This algorithm has a faster mixing rate than the Gibbs sampling algorithm \citep{geweke1991efficient}. Details of the algorithm are described in Section 1 of the online supplement \citep{GuSupplement2017}.

In the M-step, we rely on conditional maximization \citep{meng1993maximum} to update $Q(\bm{\psi}^{*} | \bm{\psi})$ in successive steps with respect to $\bm{\beta}$ and $\bm{\Sigma}$:
\begin{equation*}
\begin{split}
\hat{\bm{\beta}} &= \sum_{i=1}^N \Big{\{}  \frac{1}{G} \sum_{g=1}^G \tilde{\mathbf{z}}^{(g)}_i  \mathbf{x}_i^{\top} \Big{\}}    \Big{\{} \sum_{i=1}^N \mathbf{x}_i \mathbf{x}_i^{\top}\Big{\}} ^{-1} ,\\
\hat{\bm{\Sigma}} &=  \frac{1}{N}   \bigg{\{}    \sum_{i=1}^N  \Big{\{}   \frac{1}{G} \sum_{g=1}^G \tilde{\mathbf{z}}^{(g)}_i \tilde{\mathbf{z}}^{(g)\top}_i  \Big{\}}  -  \hat{\bm{\beta}}   \Big{\{}  \sum_{i=1}^N \mathbf{x}_i \mathbf{x}_i^{\top}  \Big{\}}    \hat{\bm{\beta}}^{\top}    \bigg{\}} ,
\end{split}
\end{equation*}
where $\{\tilde{\mathbf{z}}^{(g)}, g = 1,\ldots,G\}$ are $G$ draws from $f( \mathbf{Z} | \mathbf{Y}_{\text{obs}}, \mathbf{X}, \bm{\psi})$. To decrease the Monte Carlo errors, \citet{wei1990monte} suggested using a large $G$. 

To complete the estimation process, we derived a consistent estimator for $\bm{\gamma}$. The estimator is based on the empirical marginal distribution of the observed and imputed responses in the absence of threshold constraints \citep{olsson1979maximum}, 
\begin{equation*}
\tilde{\gamma}_{j,l} = \frac{1}{G} \sum_{g=1}^G \Phi^{-1} \Big{\{} \frac{1}{N}\sum_{i=1}^N \bm{1}\{\tilde{y}_{ij}^{(g)} \in (-\infty ,l) \} \Big{\}}, \quad l = 1, \ldots, c_j-1, j = 1,\ldots, J,
\end{equation*}
where $\tilde{y}_{ij}^{(g)} = y_{\text{obs},ij}$ if $M_i = 0$, $\tilde{y}_{ij}^{(g)} = \tilde{y}^{(g)}_{\text{imp},ij}$ and $\tilde{y}^{(g)}_{\text{imp},ij}$ is imputed through the indicator function $\bm{1}\{\tilde{z}^{(g)}_{ij} \in \Gamma_{ij} \}$ given the current estimate of $\bm{\gamma}_j$ if $M_i = 1$, and $\Phi(\cdot)$ is the cumulative distribution function of the standard Normal distribution. The estimate of $\gamma_{j,l}$ given the threshold constraints is 
\begin{equation*}
\hat{\gamma}_{j,l} = \frac{\tilde{\gamma}_{j,l} - \min \tilde{\bm{\gamma}}_j} {\max \tilde{\bm{\gamma}}_j - \min \tilde{\bm{\gamma}}_j},
\end{equation*}
where $\tilde{\bm{\gamma}}_j = (\tilde{\gamma}_{j,1}, \ldots, \tilde{\gamma}_{j,c_j-1})$. 

\subsection{Data Augmentation Algorithm}
For Bayesian inference of the MVOPT model, we assign a $\mathcal{N}(\bm{0},10^4\times\mathbf{I})$ prior distribution for $\bm{\beta}$ and a $\mathcal{IW}(J+2, (J+2)\times\mathbf{I}_{J \times J})$ prior distribution for $\bm{\Sigma}$, where $\mathcal{IW}(\nu,S_0)$ denotes the inverse-Wishart distribution with $\nu$ degrees of freedom and scale matrix $S_0$. Based on the work of \citet{albert1993bayesian}, we use a uniform prior distribution over the polytope $\mathcal{T} \subset \mathbb{R}^{c_j}$ for $\bm{\gamma}_j, j = 1,\ldots,J$. The feasible region for the parameter space of $\bm{\gamma}_j$:
\begin{equation*}
\mathcal{T} = \{ \bm{\gamma}_j = (\gamma_{j,2},\ldots,\gamma_{j,c_j-1}) \in \mathbb{R}^{c_j}: \gamma_{j,l} > \gamma_{j,l-1}, \forall \; l = 2,\ldots,c_j-1  \}.
\end{equation*}
The DA algorithm for drawing samples from the posterior distribution of $\bm{\psi}$ consists of an Imputation step that draws $\mathbf{Z}$ from $f( \mathbf{Z} | \mathbf{Y}_{\text{obs}}, \mathbf{X}, \bm{\psi})$ using the slice sampler algorithm described in Section 2.1, and three Posterior simulation (P) steps:
\begin{description}
\item P-step 1: Draw $\widetilde{\bm{\beta}} | \mathbf{Z}, \bm{\Sigma}, \mathbf{X} \sim \mathcal{N}(\bm{\mu_{\beta}}, \bm{\Sigma_{\beta}})$, where $\widetilde{\bm{\beta}} = (\bm{\beta}_1,\ldots,\bm{\beta}_J)^{\top}$, $\bm{\beta}_j$ is the $j$th row of $\bm{\beta}$, $\widetilde{\mathbf{X}}_i = \mathbf{I}_{J \times J} \otimes \mathbf{x}_i$, $\bm{\Sigma_{\beta}} = (\sum_{i=1}^N \widetilde{\mathbf{X}}_i^{\top}\bm{\Sigma}^{-1} \widetilde{\mathbf{X}}_i + 10^{-4}\times\mathbf{I}_{JP \times JP})^{-1}$ and $\bm{\mu_{\beta}} = \bm{\Sigma_{\beta}} \sum_{i=1}^N \widetilde{\mathbf{X}}_i^{\top}\bm{\Sigma}^{-1} \mathbf{z}_i$.


\item P-step 2: Draw $\bm{\Sigma} | \mathbf{Z}, \mathbf{X} \sim \mathcal{IW}(N + J + 2, \sum_{i=1}^N (\mathbf{z}_i - \bm{\beta}\mathbf{x}_i)(\mathbf{z}_i - \bm{\beta}\mathbf{x}_i)^{\top} + (J+2)\times\mathbf{I}_{J \times J})$.

\item P-step 3: Draw $\gamma_{jl}| \mathbf{z}_j, \mathbf{y}_{\text{obs},j}  \sim \mathcal{U}( \max \{ \max \{ z_{ij}:y_{ij} = l \}, \gamma_{j,l-1} \},  \min \{ \min \{ z_{ij}:y_{ij} = l+1 \}, \gamma_{j,l+1} \})$, for $l = 2, \ldots, c_j-2$, and $j = 1,\ldots, J$, where $\mathcal{U}(a,b)$ denotes the uniform distribution with support $(a,b)$. 
\end{description}
After each cycle of the algorithm, we impute the missing responses $\mathbf{Y}_{\text{mis}}$ through the indicator functions $\bm{1}\{ z_{ij} \in \Gamma_{ij} \}$, for $i = 1,\ldots, N$ and $j = 1,\ldots, J$ given the corresponding latent responses $\mathbf{Z}$ and threshold parameters $\bm{\gamma}$.

\subsection{Parameter Expansion Approach}
For the MVOPC model, we fix $\gamma_{j,1} =0$ $\forall j$ and constrain the covariance matrix $\bm{\Sigma}$ to be a correlation matrix $\mathbf{R}$. In the MCEM algorithm, the M-step with respect to $\mathbf{R}$ does not have a closed form solution \citep{chib1998analysis}, and direct maximization of the expectation of the complete-data likelihood is computationally intensive. For the Bayesian inference, the posterior distribution of $\mathbf{R}$ does not follow a known probability distribution. Thus, we use the parameter expansion (PX) technique \citep{liu1998parameter, liu1999parameter}, and propose a PX-MCEM algorithm to obtain the maximum likelihood estimates, and a PX-DA algorithm to sample from the posterior distribution of $\bm{\psi}$, respectively. These algorithms are similar to the work of \citet{zhang2012sampling}, \citet{lawrence2008bayesian} and \citet{zhang2016multiple}. 

We consider the following transformations:
\begin{equation}
\mathbf{z}_i^{*} = \mathbf{D}\mathbf{z}_i, \quad \bm{\beta}^{*} = \mathbf{D}\bm{\beta}, \quad  \mathbf{R}^{*} = \mathbf{D}\mathbf{R}\mathbf{D}, \quad  \bm{\gamma}_{j}^{*} = d_j \bm{\gamma}_{j},
\label{equation3}
\end{equation}
so that $\mathbf{R}$ is transformed into a general covariance matrix, where $\mathbf{D} = \text{diag}(d_1, \ldots, d_J)$ is a diagonal matrix with diagonal elements $d_j > 0$ $\forall j$.  The PX-MCEM and the PX-DA algorithms using the transformed parameters $(\bm{\gamma}^{*}, \bm{\beta}^{*}, \mathbf{R}^{*})$ and the latent responses $\mathbf{Z}^{*}$ proceed as the MCEM and DA algorithms described in Section 2.1 and Section 2.2, respectively. After each iteration, $(\mathbf{Z}, \bm{\gamma}, \bm{\beta}, \mathbf{R})$ are updated via the inverse transformations of identities (\ref{equation3}).


\section{Nested Multiple Imputation Procedure}
Let $Q = Q(\mathbf{Y}^{\text{mds}}_{A}, \mathbf{Y}^{\text{mds}}_{B})$ be a quantity of interest. We summarize the proposed procedure to multiply impute $(\mathbf{Y}^{\text{mds}}_{A,\text{mis}}, \mathbf{Y}^{\text{mds}}_{B,\text{mis}})$:
\begin{description}

\item Equating: Impute $\mathbf{Y}^{\text{mds}}_{A,\text{mis}}$ from the predictive distribution $f(\mathbf{Y}_{A,\text{mis}}^{\text{mds}}|\mathbf{Y}^{\text{fim}}, \mathbf{Y}_{A,\text{obs}}^{\text{mds}}, \mathbf{X})$.

\begin{description}

\item Step 1: Draw $K$ independent parameters $\tilde{\bm{\psi}}_A$ from the posterior distribution $p(\bm{\psi}_A | \mathbf{Y}^{\text{fim}}, \mathbf{Y}_{A,\text{obs}}^{\text{mds}}, \mathbf{X})$, or from the asymptotic distribution obtained by applying the EM algorithm to a bootstrapped sample of the cases.

\item Step 2: Impute $\mathbf{Y}^{\text{mds}}_{A,\text{mis}}$ through the indicator functions $\bm{1}\{\tilde{z}_{ij} \in \Gamma_{ij}\}$, where $\Gamma_{ij}$ is determined by $\tilde{\bm{\psi}}_A$ and $\tilde{z}_{ij} \sim f(z_{ij} | \mathbf{Y}^{\text{fim}}, \mathbf{Y}_{A,\text{obs}}^{\text{mds}}, \mathbf{X}, \tilde{\bm{\psi}}_A)$, $\forall i, j$.

\item Step 3: Repeat steps 1 and 2 $K$ times to create $K$ imputed datasets $\mathbf{Y}^{\text{mds},(k)}_{A,\text{imp}}$, $k = 1,\ldots, K$.
\end{description}

\item Translating: For each of the $K$ imputed datasets in Stage 1, impute $\mathbf{Y}^{\text{mds}}_{B,\text{mis}}$ from the predictive distribution $f(\mathbf{Y}_{B,\text{mis}}^{\text{mds}} |\mathbf{Y}_{A,\text{imp}}^{\text{mds}}, \mathbf{Y}_{A,\text{obs}}^{\text{oas}}, \mathbf{Y}_{B,\text{obs}}^{\text{oas}})$.

\begin{description}
\item Step 4: Draw $L$ independent parameters $\tilde{\bm{\psi}}_B^{(k)}$ from the posterior distribution $p(\bm{\psi}_B | \mathbf{Y}^{\text{oas}}_{A,\text{obs}}, \mathbf{Y}^{\text{mds},(k)}_{A,\text{imp}})$, or from the asymptotic distribution obtained by applying the EM algorithm to a bootstrapped sample of the cases.

\item Step 5: Impute $\mathbf{Y}^{\text{mds}}_{B,\text{mis}}$ through the indicator functions $\bm{1}\{\tilde{z}_{ij} \in \Gamma_{ij}\}$, where $\Gamma_{ij}$ is determined by $\tilde{\bm{\psi}}_B^{(k)}$ and $\tilde{z}_{ij} \sim f(z_{ij} | \mathbf{Y}^{\text{mds},(k)}_{A,\text{imp}}, \mathbf{Y}^{\text{oas}}_{A,\text{obs}}, \mathbf{Y}^{\text{oas}}_{B,\text{obs}}, \tilde{\bm{\psi}}_B^{(k)})$, $\forall i, j$.

\item Step 6: Repeat steps 4 and 5 $L$ times to create $L$ imputed datasets $\mathbf{Y}_{B,\text{imp}}^{\text{mds},(k,l)}$, $l = 1,\ldots,L$.
\end{description}

\item Combining Rules: The estimate of $Q$ and its sampling variance are $\hat{Q}^{(k,l)} = \hat{Q}^{(k,l)}(\mathbf{Y}_{\text{com}}^{(k,l)})$ and $U^{(k,l)} = U^{(k,l)}(\mathbf{Y}_{\text{com}}^{(k,l)})$ respectively, where each of the complete datasets $\mathbf{Y}_{\text{com}}^{(k,l)} = (\mathbf{Y}^{\text{mds}}_{A,\text{obs}}, \mathbf{Y}^{\text{mds}}_{B,\text{obs}}, \mathbf{Y}_{A,\text{imp}}^{\text{mds},(k)}, \mathbf{Y}_{B,\text{imp}}^{\text{mds},(k,l)})$, for $k = 1,\ldots,K$, and $l = 1,\ldots, L$. The overall estimate of $Q$ and its sampling variance are obtained using the nested multiple imputation combining rule, confidence intervals and significance tests are based on a Student-$t$ reference distribution \citep{shen2000nested, harel2003strategies, rubin2003nested}.


\end{description}

\section{Simulation Study}
We examined the performance of the MVOP model in comparison to existing methods for imputing incomplete multivariate ordinal variables using a simulation study. 

\subsection{Partially Simulated Data}
The simulation study was based on observed FIM assessments and MDS assessments on admission for patients in SNFs, and missing MDS assessments were artificially generated. To generate incomplete data sets, we fitted a logistic regression model to the entire dataset where the explanatory variables comprised $\mathbf{Y}^{\text{fim}}$, patients' age and patients' gender,
\begin{equation}
\text{logit}\{ \text{Pr}(M_i = 1| \mathbf{y}_{i}^{\text{fim}}, \mathbf{x}_i) \} = \alpha_0 + \sum_{j=1}^{J_1}\alpha_{j}y_{ij}^{\text{fim}} + \alpha_{J_1+1}x_{i1} + \alpha_{J_1+2}x_{i2},
\end{equation}
where $\mathbf{y}^{\text{fim}}_i = (y_{i1}^{\text{fim}},\ldots,y_{iJ_1}^{\text{fim}})$, $\mathbf{x}_i = (x_{i1}, x_{i2})$, and $x_{i1}$ and $x_{i2}$ denote the age and gender of patient $i$, respectively. This resulted in estimated regression coefficients $\hat{\bm{\alpha}}^{'} = (\hat{\alpha}_0, \hat{\bm{\alpha}}_1^{'})$, where $\hat{\alpha}_0$ is the estimated intercept and $\hat{\bm{\alpha}}_1 = (\hat{\alpha_1}, \ldots, \hat{\alpha}_{J_1+2})$ is a vector of estimated regression coefficients for the other predictors. A simple random sample of $n$ = 1,000 patients was then drawn from the set of patients in SNFs, and $M_i$ ($i = 1,\ldots,n$) was sampled from a Bernoulli distribution with probability $\text{Pr}(M_i = 1|\mathbf{y}_{i}^{\text{fim}}, \mathbf{x}_i) = F(\tilde{\alpha}_0 + \sum_{j=1}^{J_1}\hat{\alpha}_{j}y_{ij}^{\text{fim}} + \hat{\alpha}_{J_1+1}x_{i1} + \hat{\alpha}_{J_1+2}x_{i2})$, where $F(\cdot)$ is the c.d.f. of a specified distribution and $F^{-1}(\cdot)$ is the link function \citep{mccullagh1989generalized}. We considered three choices of $F(\cdot)$, the logistic distribution, the Cauchy distribution, which is symmetric but has heavier tails than the logistic distribution, and the Box-Cox family distributions \citep{guerrero1982use}. The Box-Cox distribution takes the form
\[ 
F_{\lambda}(x) = \begin{cases}
0, \hspace{1.7cm}\;\; x < -\frac{1}{\lambda}, \lambda > 0\\
\frac{(1+\lambda x)^{1/\lambda}}{(1+\lambda x)^{1/\lambda}+1}, \;\; 1+\lambda x > 0, \lambda \neq 0 \\
\frac{\exp(x)}{1 + \exp(x)}, \hspace{0.5cm}\;\; \lambda = 0\\
1, \hspace{1.7cm}\;\; x > -\frac{1}{\lambda}, \lambda < 0
\end{cases}.
\]
This distribution allows us to assess the effect of skewness in the missing data mechanism. It is positively skewed for $\lambda > 0$ and negatively skewed for $\lambda < 0$; here, $\lambda$ was fixed at either -0.3 or 0.3. The value of $\tilde{\alpha}_0$ was fixed so that $p_{\text{mis}} = n_1/n$ is either $20\%$, $40\%$ or $60\%$, where $n_1$ is the number of patients who have missing assessments. MDS assessments for patient $i$ were deleted to create incomplete data set when $M_i = 1$. For each configuration, 1,000 replications were produced. 

The methods examined in the simulations were IPSM, LVM and the MVOP models implemented using both EM and DA algorithms: MVOPT-DA, MVOPT-EM, MVOPC-DA and MVOPC-EM. For IPSM, we estimated the propensity score using the logistic regression model: $\text{logit}\{e(\mathbf{y}^{\text{fim}}_i, \mathbf{x}_i)\} = \xi_0 + \sum_{j=1}^{J_1} \xi_j y^{\text{fim}}_{ij} + \xi_{J_1+1} x_{i1} + \xi_{J_1+2} x_{i2}$. For both IPSM and LVM, we used the nearest-neighbor matching algorithm to find a potential donor. Ten multiple imputations were generated using each of the methods.

We examined the performance of the different methods on two estimands: (1) the population mean total score of items in MDS, $Q_1 \equiv E(S^{\text{mds}})$, where $S_{i}^{\text{mds}} = \sum_{j}y_{ij}^{\text{mds}}$, $i = 1,\ldots,n$; (2) the pairwise Goodman and Kruskal's $\gamma$ rank correlation coefficients \citep{goodman1954measures} between $J_{\text{mds}}$ items in MDS, $Q_2 \equiv \{\gamma(\mathbf{y}^{\text{mds}}_{j}, \mathbf{y}^{\text{mds}}_{k}), 1 \leq j < k \leq J_{\text{mds}} \}$. For each method, at each configuration, and at each of the 1,000 replications, we recorded $\hat{Q}_m$, $m = 1,2$, their estimated sampling variances, the corresponding root mean square errors (RMSEs), the $95\%$ interval estimate widths, and determined whether the intervals covered or did not cover the true value. Using these values, we calculated for each approach and each configuration, the average coverage rate, the bias, the mean estimated sampling variance, the mean RMSE, and the mean interval width. Because the simulations are based on 1,000 replicates for each configuration, observed coverage of 93.7$\%$ or above is not statistically distinguishable from the nominal
level. In addition, we view observed coverage of 90$\%$ as indicative of a modest deficit in coverage. 

For each configuration, we also calculated a loss function based on the negatively oriented interval scores \citep[Equation (61)]{gneiting2007strictly}. This loss function provides flexible assessment of coverage by accounting for the distance between the interval estimate and the estimand. For estimand $Q_{m}$, the loss function for interval estimate $I$, has the form 
\begin{equation}
\lambda(I) + \frac{2}{\alpha} \text{inf}_{\eta \in I}|Q_{m}-\eta|,
\end{equation} 
where $\alpha = 0.05$ and $\lambda(I)$ denotes the Lebesgue measure of the interval estimate $I$.


The simulations were implemented using R 3.1.0 \citep{R-Core-Team:2014aa}. The proposed EM and DA algorithms were implemented in C++ for efficiency. For the EM algorithms, we generated $G = 100$ samples from $f( \mathbf{Z} | \mathbf{Y}_{\text{obs}}, \mathbf{X}, \bm{\psi})$ in the E-step, and calculated the observed-data likelihood using a Monte Carlo method \citep{genz1992numerical} to monitor the convergence of the MCEM algorithm. For the DA algorithms, multiple parallel chains of 50,000 iterations with dispersed initial values were generated. Standard MCMC convergence diagnostics such as Gelman-Rubin Statistic \citep{gelman1992inference}, trace plots, and autocorrelation plots were examined for a small sample of the simulations, and did not indicate failure to converge.

\subsection{Results}

Table~\ref{table_s1} displays the mean biases, variances, RMSEs, coverages, interval widths and interval estimate loss function of the population mean total score of items in MDS, $Q_1$, for configurations, where $F(\cdot)$ is the logistic distribution and $p_{\text{mis}} = \{0.2, 0.4, 0.6\}$. Although some methods show modest deficits in coverage in some scenarios, all of the methods yield coverage that is generally either at or above the nominal level, statistically indistinguishable from the nominal level, or indicative of only a modest deficit in coverage. Compared to all of the methods that were examined, MVOP models implemented using the DA algorithms have coverages that are closest to nominal across all configurations. IPSM has coverages that are slightly smaller than LVM for $p_{\text{mis}}={0.2,0.4}$, and worse than LVM for $p_{\text{mis}}=0.6$. When $p_{\text{mis}}=0.6$, the parametric models underlying LVM impute the missing values with less bias than the propensity score model used in IPSM. Similar results were observed when predictive mean matching was compared to IPSM \citep{andridge2010review}. The MVOP models implemented using the DA algorithms generally have the smallest biases and RMSEs, while the MVOP models implemented using the EM algorithms and the bootstrap method have the largest biases, variances, RMSEs and interval widths. Because some methods have lower coverage but with shorter intervals, and some have higher coverage with wider intervals, we used the loss function in Equation (4.2) to compare the methods on coverage and interval width simultaneously. Generally, the MVOPT models implemented using the DA algorithms have the smallest interval score loss followed by LVM.

\begin{table}
\centering
\begin{threeparttable}
\caption{\footnotesize{Biases, variances, RMSEs, $95\%$ interval coverages, $95\%$ confidence interval widths, and interval estimate loss function (Equation (4.2)) for the population mean total score of items in MDS, $Q_1$, given that $n = 1000$ and $F(\cdot)$ is the logistic distribution.}}
\label{table_s1}
\begin{tabular}{clrccccc }
    \hline
$p_{mis}$  &Method & Bias & Variance & RMSE & Coverage & Width & Equation (4.2)\\ 
  \hline
&IPSM\tnote{a} &-0.013 & 0.014 & 0.124 & 93.2 & 0.460 & 0.607\\ 
&LVM\tnote{b} & -0.047 & 0.014 & 0.124 & 93.9 & 0.459 & 0.588\\
0.2&MVOPT\tnote{c}-DA\tnote{d} & -0.015 & 0.014 & 0.116 & 95.5 & 0.472 & 0.547 \\
&MVOPC\tnote{e}-DA & -0.012 & 0.013 & 0.115 & 94.3 & 0.453 & 0.569\\ 
&MVOPT-EM\tnote{f} & -0.017 & 0.029 & 0.133 & 98.0 & 0.735 & 0.678\\  
&MVOPC-EM & -0.032 & 0.030 & 0.130 & 97.8 & 0.752 & 0.677\\ 
  \hline
&IPSM & -0.015 & 0.016 & 0.147 & 90.4 & 0.500 & 0.752 \\  
&LVM & -0.070 & 0.020 & 0.159 & 91.8 & 0.565 & 0.767\\ 
0.4&MVOPT-DA & -0.030 & 0.025 & 0.139 & 96.1 & 0.641 & 0.715 \\ 
&MVOPC-DA & -0.025 & 0.020 & 0.141 & 94.8 & 0.564 & 0.793\\ 
&MVOPT-EM & -0.070 & 0.039 & 0.175 & 97.3 & 0.885 & 0.823\\
&MVOPC-EM & -0.073 & 0.039 & 0.166 & 97.2 & 0.886 & 0.836\\ 
  \hline
&IPSM & 0.062 & 0.034 & 0.228 & 89.6 & 0.748 & 1.191\\ 
&LVM & -0.050 & 0.035 & 0.196 & 92.2 & 0.757 & 1.007\\ 
0.6&MVOPT-DA & -0.027 & 0.050 & 0.183 & 97.9 & 0.931 & 0.984\\ 
&MVOPC-DA & -0.007 & 0.033 & 0.179 & 95.3 & 0.749 & 0.871\\
&MVOPT-EM & -0.180 & 0.061 & 0.269 & 91.7 & 1.137 & 1.084\\
&MVOPC-EM & -0.167 & 0.063 & 0.263 & 92.4 & 1.164 & 1.024\\ 
\hline
\end{tabular}
\begin{tablenotes}
\item[a] IPSM: imputation by propensity score matching;
\item[b] LVM: latent variable matching;
\item[c] MVOPT: multivariate ordinal probit model with threshold constraints;
\item[d] DA: data augmentation algorithm;
\item[e] MVOPC: multivariate ordinal probit model with correlation constraints;
\item[f] EM: expectation-maximization algorithm.
\end{tablenotes}
\end{threeparttable}
\end{table}

\begin{figure}
\vspace{6pc}
\subfigure[Bias]{
\includegraphics[scale=0.40]{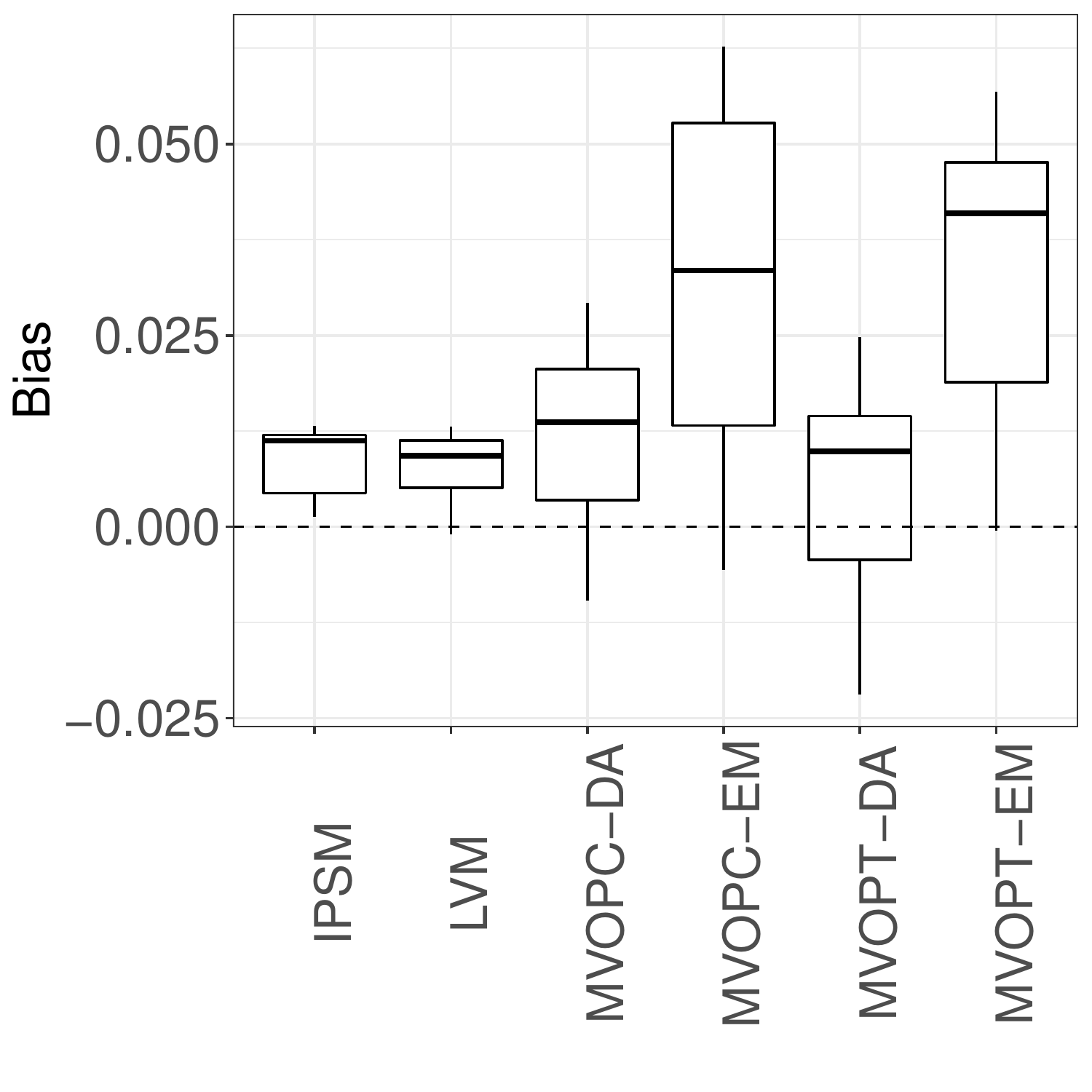}}
\subfigure[Coverage]{
\includegraphics[scale=0.40]{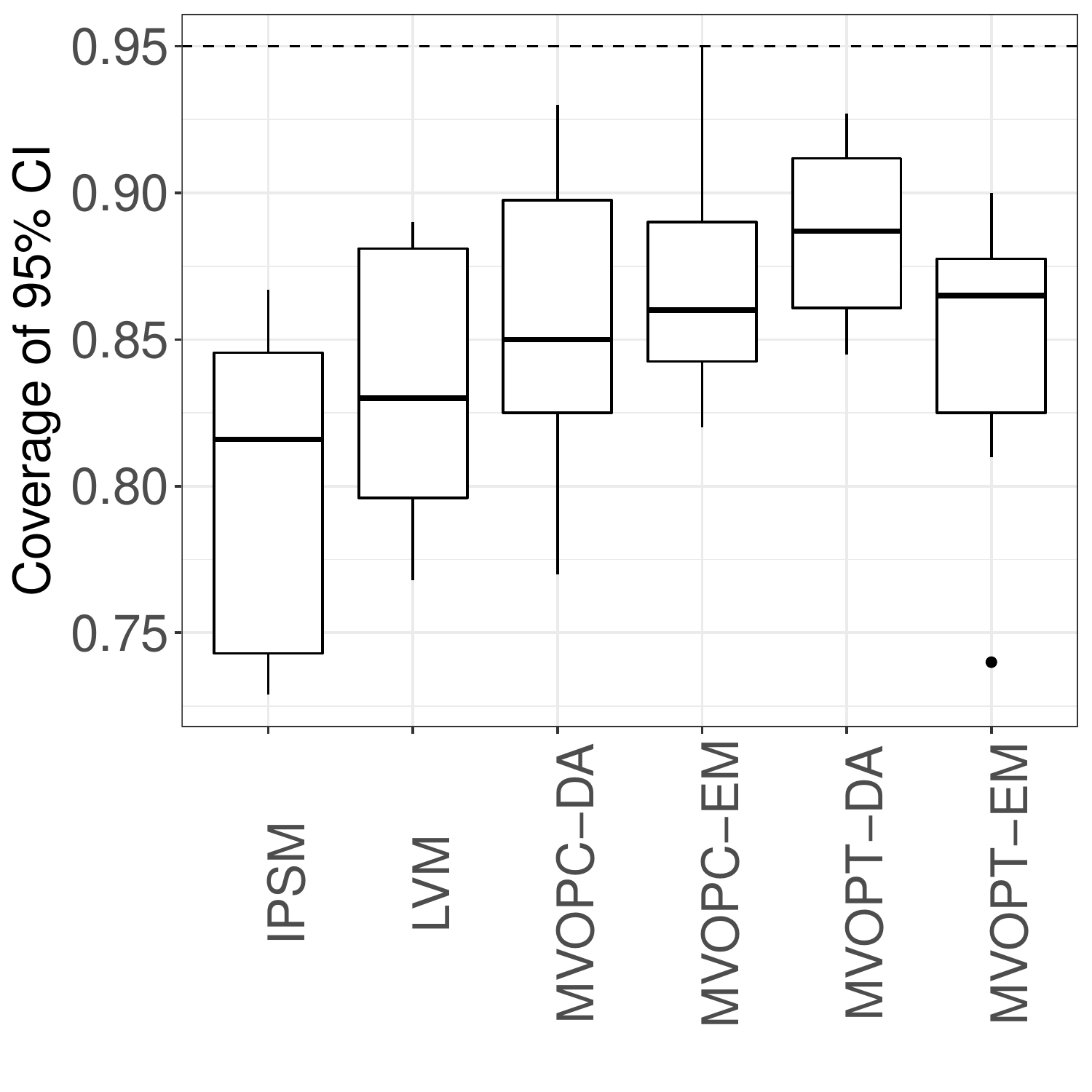}}\\
\subfigure[CI Width]{
\includegraphics[scale=0.40]{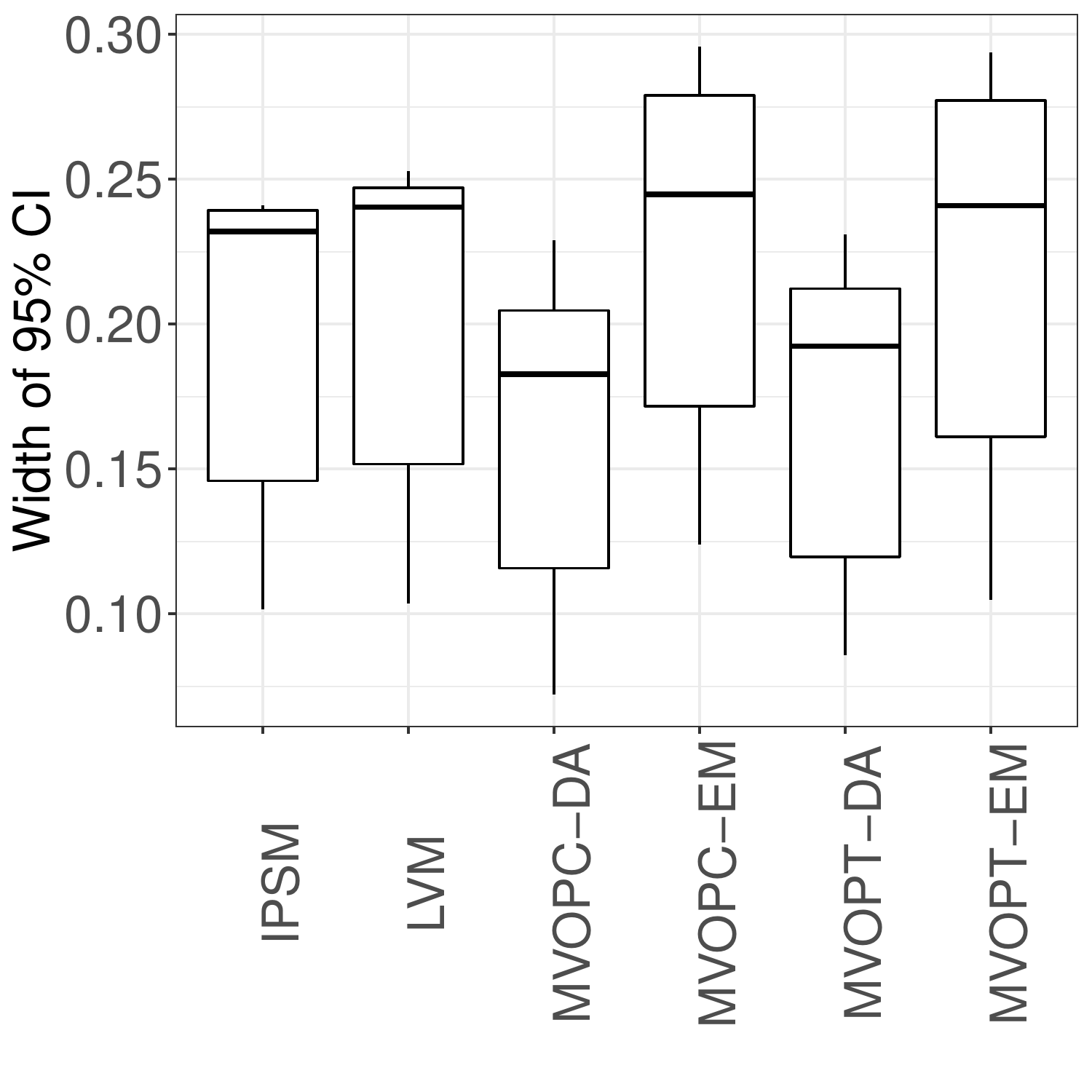}}
\subfigure[Interval Score Loss]{
\includegraphics[scale=0.40]{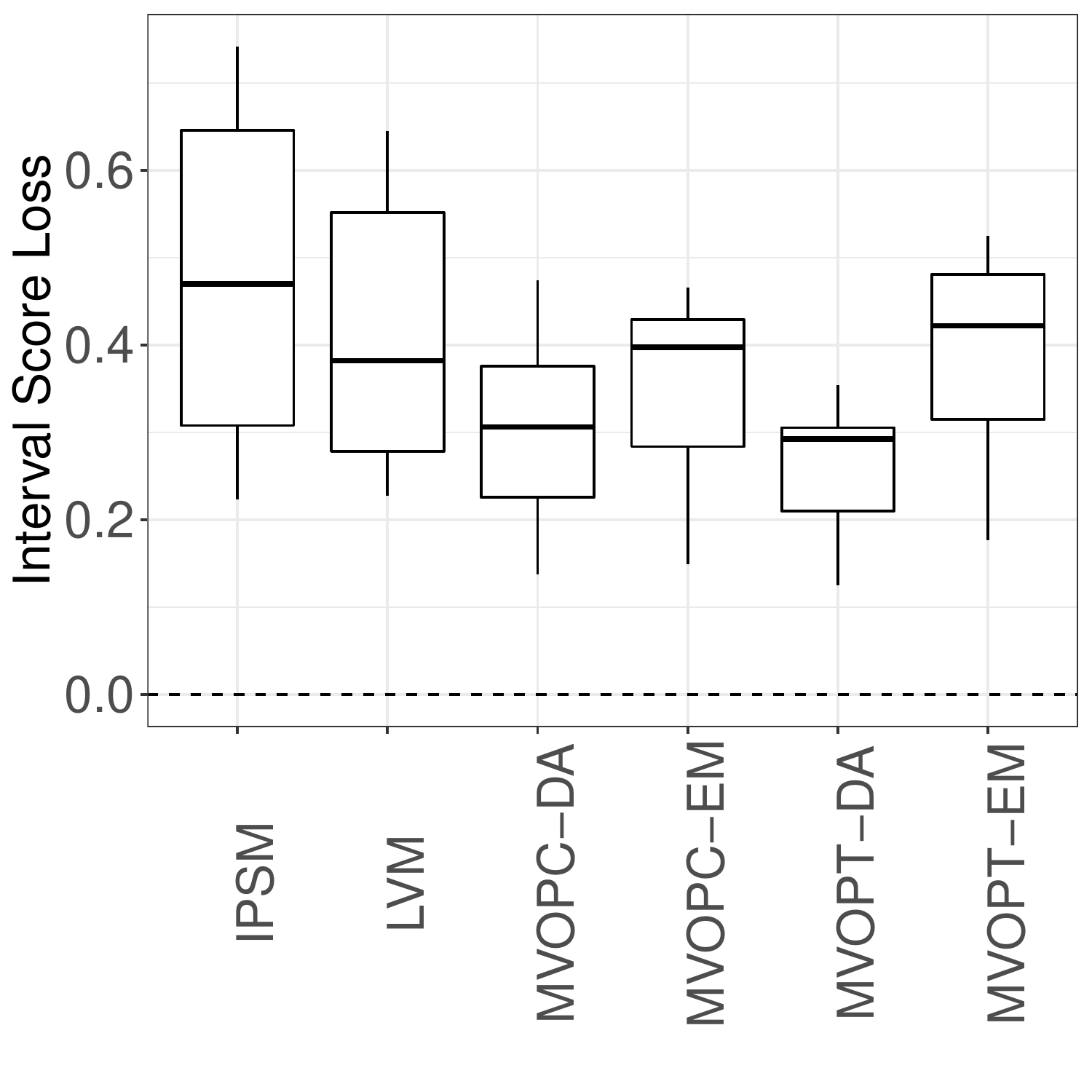}}
\caption{Distribution across 1,000 simulation replications of (a) biases, (b) coverages of $95\%$ confidence interval, (c) widths of $95\%$ confidence interval and (d) interval score loss for the pairwise rank correlation coefficients between items in MDS, $Q_2$, given that $F(\cdot)$ is the logistic distribution and $p_{\text{mis}} = 0.6$.}
\label{logit_figure3}
\end{figure}

Figure~\ref{logit_figure3} displays the distribution of biases, $95\%$ interval coverages, interval widths and interval score loss of the pairwise rank correlation coefficients between items in MDS, $Q_2$, for configurations where $F(\cdot)$ is the logistic distribution and $p_{\text{mis}} =0.6$. The MVOP models except for MVOPC-DA have coverages that are close to nominal, while IPSM and LVM have median coverage that is lower than $85\%$. However, except for MVOPT-DA, the other MVOP models have biases that are larger than IPSM and LVM. As with $Q_1$, MVOPT-EM and MVOPC-EM have the largest biases and interval lengths, but their coverages are closer to nominal when compared to LVM and IPSM. Lastly, MVOPT-DA has better coverages and smaller interval score loss than MVOPC-DA. These trends are similar to the ones observed with $p_{\text{mis}} = \{0.2, 0.4\}$ (see Figures 5 - 6 in Section 2 of the online supplement \citep{GuSupplement2017}).

Because MVOPT-DA generally has the best operating characteristics when $F(\cdot)$ is the logistic distribution, we only include this method when examining the effects of propensity score model misspecification. Table 1 in Section 2 of the online supplement \citep{GuSupplement2017} displays the results for the population mean total score of items in MDS when $F(\cdot)$ is the Cauchy distribution or the Box-Cox family distribution with $\lambda = \{-0.3, 0.3\}$. The performance of IPSM is sensitive to misspecification of the propensity score model. For example, when $p_{\text{mis}} = 0.6$ and $F(\cdot)$ is the Box-Cox family distribution with $\lambda = -0.3$, the coverage of IPSM is only $82\%$ and its interval score loss is larger than LVM and MVOPT-DA. In contrast, LVM and MVOPT-DA are robust to different link functions. MVOPT-DA has better coverages and smaller biases than LVM across all of the configurations that were examined, and generally has smaller interval score loss than LVM. Figures 7 - 15 in Section 2 of the online supplement \citep{GuSupplement2017} display the results for the pairwise rank correlation coefficients between items in MDS when $F(\cdot)$ is the Cauchy distribution and the Box-Cox family distribution with $\lambda = \{-0.3, 0.3\}$, respectively. IPSM, LVM and MVOPT-DA have similar point estimates, but MVOPT-DA has better coverages and smaller interval score loss than LVM and IPSM in most of the examined configurations. IPSM has the lowest coverages, and the median of its coverages is about $72\%$ when $F(\cdot)$ is the Box-Cox distribution with $\lambda = -0.3$ and $p_{\text{mis}} = 0.6$. When the percentage of missingness decreases, the coverages of MVOPT-DA are closer to nominal. 

\subsection{Sensitivity of the Methods to the Conditional Independence and MAR Assumptions}

The proposed methods rely on the validity of the conditional independence and MAR assumptions (Section 1.3). We conducted an additional simulation study to examine the plausibility of these assumptions in this analysis.

One clinical variable that is recorded for patients in IRFs is their swallowing status at discharge. Swallowing status is a categorical variable with three categories: $``$Regular Food$"$, $``$Modified Food Consistency/Supervision$"$ and $``$Tube/Parenteral$"$. Swallowing status is correlated with patients' self-care functional status as well as patients' discharge destination. We recoded the swallowing status using two dummy variables, which were added to the Equation (4.1). The setup of the coefficients in Equation (4.1) was similar to the one in Section 4.1, and we also considered the three different link functions and three possible values for $p_{\text{mis}}$. Because MVOPT-DA has the best operating characteristics, we only examined the validity of the conditional independence and MAR assumptions with this model. Swallowing status was not included when fitting the MVOPT-DA model to address the possibility that physicians may have used unobserved clinical information when selecting between the two possible discharge destinations. The fitted MVOP model potentially violates both the conditional independence and the MAR assumptions.

Table 2 and Figure 16 - 19 in Section 2 of the online supplement \citep{GuSupplement2017} display the results for the population mean total score of items in MDS and for the pairwise rank correlation coefficients between items in MDS, respectively. MVOPT-DA generally provides statistically valid inferences. When the percentage of missingness decreases, the biases, variances, RMSEs and interval widths of MVOPT-DA decrease.


\section{Motivating Example Revisited}
\subsection{Data}
FIM, MDS and OASIS include similar functional status items, but they have differences in the rating levels (i.e., $``$independence$"$ is reflected by a higher score in FIM but a lower score in MDS). To increase the consistency of the items in these three instruments, we reversed the rating levels of FIM prior to the analysis such that in all three instruments lower rating levels represent better functional status. In addition, we recoded any MDS items with score of 7 or 8 (activity occurred only once or twice or activity did not occur) as a score of 4 (totally dependent) \citep{wysocki2015functional}. We also combined the scores 3, 4, and 5 in the item $``$Feeding or Eating$"$ in OASIS due to a small proportion ($<$ 1$\%$) of patients responding at these levels. After recoding, the items in FIM, MDS and OASIS have seven, five and four rating levels, respectively, except for the item $``$bathing$"$ in OASIS that has seven levels.

Patients' demographic characteristics are summarized in Table~\ref{demographic}. Table~\ref{table22} displays patients' functional assessments in the three instruments. Patients who were discharged home have an average FIM total score of 17.19 ($SD$ = 6.21), while the average of the total score for patients who were discharged to SNFs is 27.41 ($SD$ = 7.46). This suggests that patients that were released home have better functional status when they were discharged from IRFs. Table~\ref{table22} also shows that patients who were either released home or to SNFs have smaller average total scores at the later assessment date, suggesting that the functional status for most of patients improves over the course of their post-acute stay. The magnitude of improvement among the subsample of patients who received home health care appears to be larger than those who stayed in SNFs. $84.5\%$ of the patients who recived home health improve their functional status, while only $48.2\%$ of the patients in SNFs.

\begin{table}
\centering
\caption{Summary of the observed covariates for patients.}
\label{demographic}
\begin{tabular}{lccc}
    \hline
Variable & SNF & Home Health & Overall \\ 
  \hline
Age   & 77.17 (9.62) & 76.40 (10.05) & 76.81 (9.83) \\
Gender, female ($\%$) & 53.0 & 53.2 & 53.1 \\
Race, white ($\%$) &  81.2 & 77.0 & 79.2 \\
Marital status, married ($\%$) & 42.2 & 50.0  & 45.9 \\
   \hline
\end{tabular}
\end{table}

\begin{table}
\centering
\begin{threeparttable}
\caption{Summary of patient's functional outcomes in three instruments.}
\label{table22}
\begin{tabular}{ccccc}
    \hline
Instrument &Variable & SNF & Home Health & Overall \\ 
  \hline
FIM  & Score &  27.41(7.46) & 17.19(6.21) & 22.63(8.59) \\
  \hline
   & Score\tnote{a} at time 1 & 18.38 (2.85) & -&- \\
MDS &Score at time 2 & 17.43 (3.66) &- &- \\
&Difference\tnote{b} & -0.95 (2.41) &- &- \\
&Improved\tnote{c} ($\%$) & 48.2  & - & -\\
  \hline
 & Score at time 1 &-& 15.60 (4.12) &- \\  
 OASIS & Score at time 2 &-& 10.59 (4.75)  &- \\  
 &Difference &- &-5.01 (4.07)  &- \\
&Improved ($\%$) &-& 84.5   &- \\
   \hline
\end{tabular}
\begin{tablenotes}
\item[a] Score: the total score of functional assessments in each instrument;
\item[b] Difference: the difference in total scores measured on admission and at a later assessment date;
\item[c] Improved: the proportion of patients who experience functional improvement.
\end{tablenotes}
\end{threeparttable}
\end{table}



\subsection{Imputation Model}
We illustrate the proposed nested multiple imputation procedure using the complete data set of 72,575 patients. In the first imputation stage, we impute the unmeasured assessments in MDS using the MVOPT-DA method described in Section 2.2. Age, gender, race and marital status are included in the model. Ten parallel chains of 50,000 iterations with dispersed initial values are generated, resulting in ten imputed data sets. 

In the Translating step, we consider two possible models to illustrate the flexibility of the proposed procedure for translating assessments without re-equating the instruments. The first model is a linear regression model, $E(s_{1} | s_{2}) = \xi_0 + \xi_1 s_{2}$, where $s_{1}$ and $s_{2}$ denote the total scores of the imputed and observed items in MDS and OASIS on admission, respectively. The unmeasured total scores in MDS at the later assessment date are imputed using the estimates of $\xi_0$ and $\xi_1$ and the observed total scores in OASIS at the later assessment date. The second model is the MVOPT model, which models the joint distribution of all individual items in the imputed MDS and the observed OASIS instruments, $f(\mathbf{Y}_{A,\text{imp}}^{\text{mds}}, \mathbf{Y}_{A,\text{obs}}^{\text{oas}} | \bm{\psi})$. The unmeasured individual items in MDS at the later assessment date are imputed using the estimates of $\bm{\psi}$ and the observed individual items in OASIS at the later assessment date, $\mathbf{Y}_{B,\text{obs}}^{\text{oas}}$. For the MVOPT model, the DA algorithm in Section 2.2 is used to generate multiple imputations. Ten imputed data sets are generated in the second stage, resulting in 100 complete data sets.

We also examine the conversion table method and LVM to equate the MDS and OASIS instruments in the Equating step, and the linear regression model to impute the missing total scores in MDS in the Translating step. For LVM, in order to accommodate patients' covariates, we first partition the sample into five subclasses by sub-classifying at the quintiles of the distributions of the estimated propensity scores, $\widehat{\text{Pr}}(M_i = 1 | \mathbf{x}_i)$, and then impute the unmeasured assessments within each subclass.

\subsection{Model Diagnostics}
\begin{figure}
\vspace{6pc}
\includegraphics[scale=0.35]{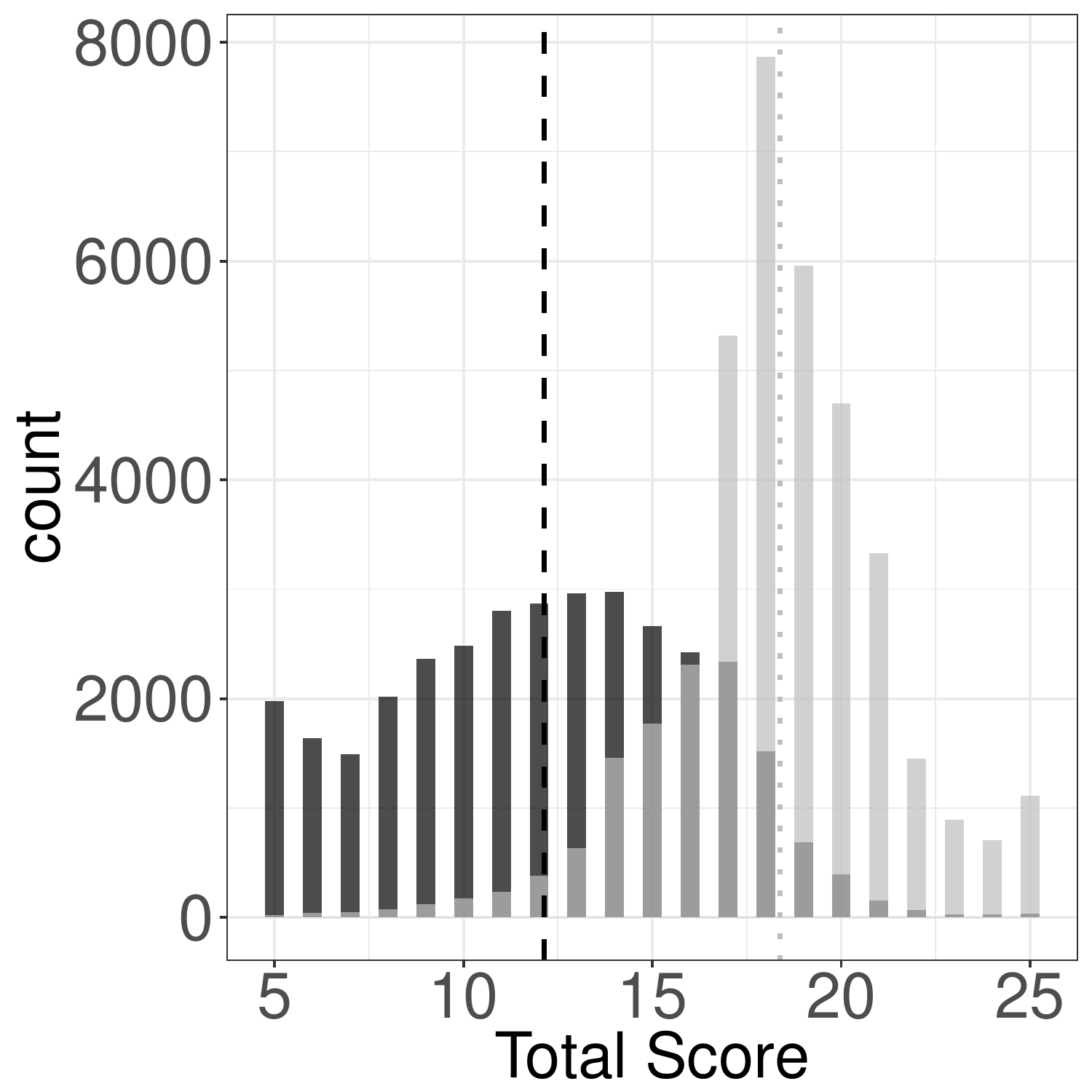}
\caption{Histograms of the observed (gray) and imputed (black) total scores in MDS. The gray dotted line and black dashed line are the average observed and imputed total scores, respectively.}
\label{totalscore1}
\end{figure}

As suggested by \citet{gelman2005multiple} and \citet{abayomi2008diagnostics}, we evaluated the imputation model by comparing the distributions of the observed and the imputed values. Patients who were released home have smaller total MDS scores (see Figure~\ref{totalscore1}), and are more likely to be at lower levels of each item (not shown). These patterns are consistent with the patterns that are observed in FIM.

We further examined the imputation model using posterior predictive checks \citep{gelman1996posterior, burgette2010multiple, he2012diagnosing, si2013nonparametric, si2016bayesian}. We first created $S = 1,000$ complete data sets $D^{(s)} = (\mathbf{Y}_{\text{obs}}, \mathbf{Y}_{\text{mis}}^{(s)})$ $(s = 1,\ldots,S)$ and replicated data sets $R^{(s)} = \mathbf{Y}_{\text{rep}}^{(s)}$ in which both $\mathbf{Y}_{\text{obs}}$ and $\mathbf{Y}_{\text{mis}}$ are simulated from the imputation model. We then compared each $D^{(s)}$ with its corresponding $R^{(s)}$ on three test statistics in the first stage imputation: (1) the mean total score of items in MDS, $T_1 \equiv \sum_{i,j} y_{ij}^{\text{mds}} / N$; (2) the proportion of response levels in each of the $J_{\text{mds}}$  items in MDS, $T_2 \equiv \{n_{lj} / N, l = 1,\ldots,c_j, j = 1,\ldots,J_{\text{mds}}\}$, where $n_{lj}$ is the number of responses at level $l$ in item $j$; and (3) the pairwise Goodman and Kruskal's $\gamma$ rank correlation coefficients between items in MDS, $T_3 \equiv \{\gamma(\mathbf{y}_{j}, \mathbf{y}_{k}), 1 \leq j < k \leq J_{\text{mds}} \}$. Let $T_{m,D^{(s)}}$ and $T_{m,R^{(s)}}$, $m =1,2,3$, be the values of $T_{m}$ computed with $D^{(s)}$ and $R^{(s)}$, respectively. For each $T_m$ $(m = 1,2,3)$, we computed the two-sided posterior predictive probability ($ppp$),
\begin{equation*}
ppp_m = (2 / S) \times \min \Big{(}  \sum_{s=1}^S \mathbf{1}(T_{m,D^{(s)}} > T_{m,R^{(s)}}), \sum_{s=1}^S \mathbf{1}(T_{m,D^{(s)}} < T_{m,R^{(s)}})   \Big{)}, 
\end{equation*}
where $\mathbf{1}(\cdot)$ is the indicator function that is equal to 1 if the condition is satisfied and 0 otherwise. A small $ppp$ indicates that $T_{D^{(s)}}$ and $T_{R^{(s)}}$ deviate from each other in one direction, which suggests that the imputation model distorts the data characteristics captured by $T_m$.

\begin{figure}
\vspace{6pc}
\subfigure{
\includegraphics[scale=0.3]{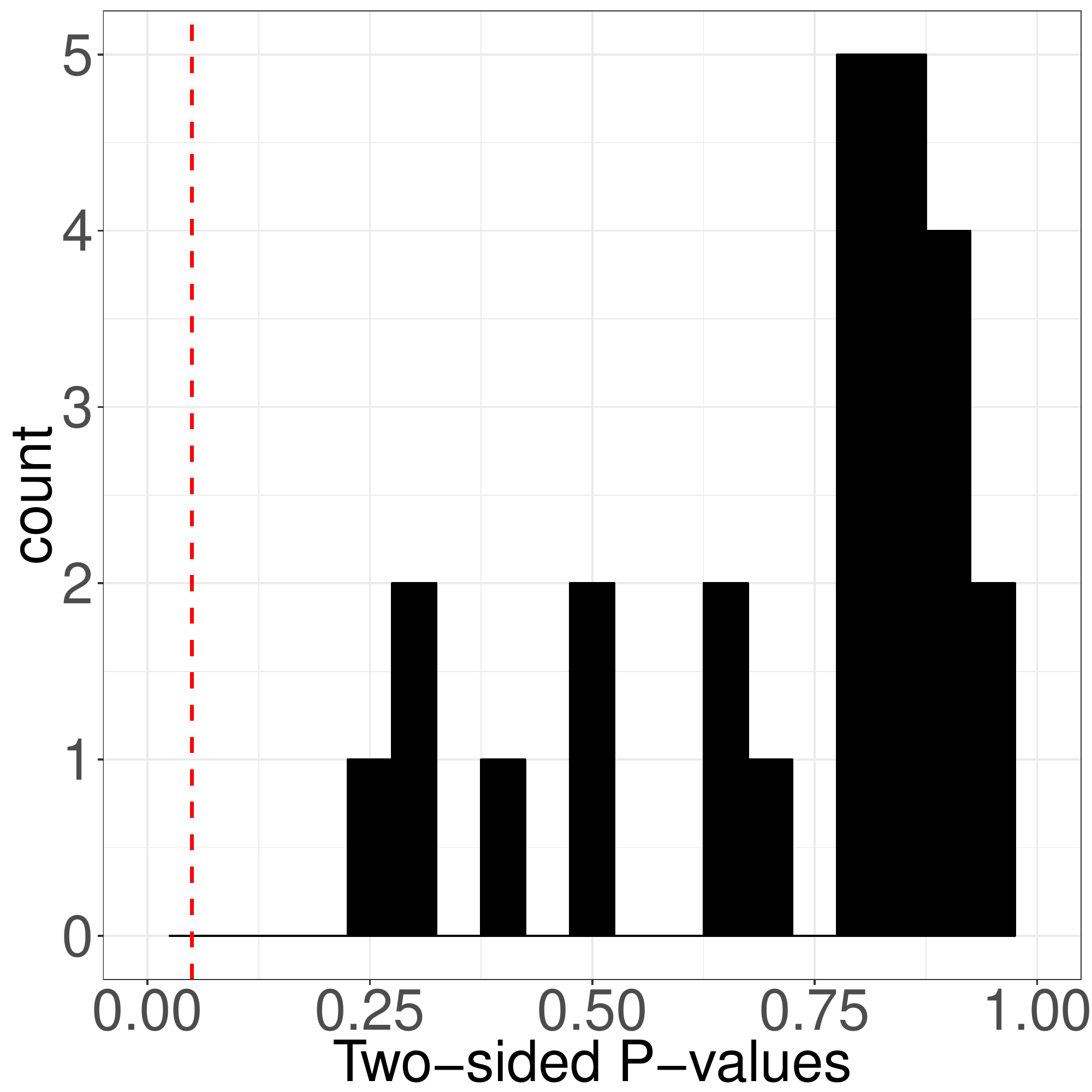}}
\subfigure{
\includegraphics[scale=0.3]{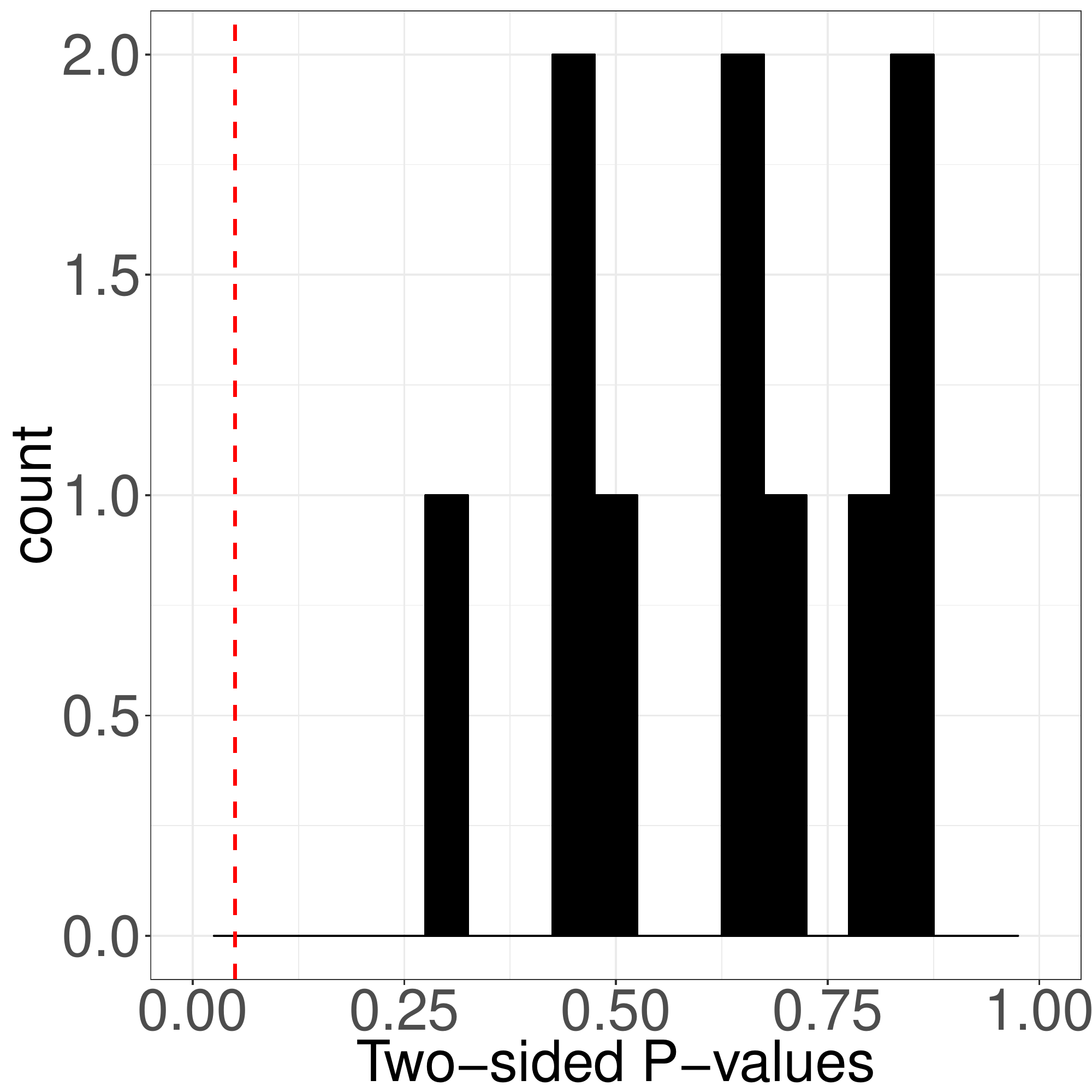}}
\caption{Histograms of the two-sided posterior predictive probabilities ($ppp$) for $T_2$ (left panel) and $T_3$ (right panel). The red dashed line corresponds to a threshold value of 0.05.}
\label{ppc}
\end{figure}

To obtain the pairs $(D^{(s)}, R^{(s)})$, we added a step to the DA algorithm that replaced all the values of $\mathbf{Y}_{obs}$ and $\mathbf{Y}_{mis}$ using the sampled parameter values at each iteration. We calculated the test statistics $T_1$ based on $1,000$ complete and replicated data sets, and their differences $T_{1,D^{(s)}} - T_{1,R^{(s)}}$, $s = 1,\ldots,1,000$. The estimated two-sided $ppp_1 = 0.446$, which does not indicate a deficiency in the imputation model for $T_1$. The left and right panel of Figure~\ref{ppc} show the histogram of the two-sided $ppp$ values for $T_2$ and $T_3$, respectively. None of the $ppp_2$ and $ppp_3$ values are below 0.05. Thus, we do not observe implausible imputations. Similar model diagnostics were performed for the second stage imputation, and no significant lack of model fit was detected (not shown).

Because posterior predictive checks may not be well calibrated \citep{hjort2006post}, we also examined the imputation performance using a sample partitioning method. Patients in SNFs were partitioned into a training sample that included 90$\%$ of the patients, and the remaining 10$\%$ served as a test sample. We fit the MVOPT model to the training sample and predicted the assessments of the test sample. We repeated this partitioning and prediction process ten times, and in each replication we compared the distributions of the total mean score and the pairwise rank correlation coefficients of the predicted MDS assessments to the observed ones. Table 3 of the online supplement \citep{GuSupplement2017} shows the results of the ten replications. No significant lack of model fit is detected.

\subsection{Analysis of Multiply Imputed Data}
We compared the rates of functional change experienced by patients treated in SNFs and those treated by home health agencies using the observed and imputed assessments in MDS. 

We define $d_{\text{snf}}$ and $d_{\text{hh}}$ to be the average change in total scores of the items in MDS over two assessments after discharge from IRFs for patients treated in SNFs and by home health agencies, respectively: 
\begin{equation*}
d_{\text{snf}} =  \frac{1}{N_1} \sum_{i=1}^{N_1} S^{\text{snf}}_{2,i} -  \frac{1}{N_1} \sum_{i=1}^{N_1} S^{\text{snf}}_{1,i}, \quad \text{and} \quad
d_{\text{hh}} = \frac{1}{N_2} \sum_{i=1}^{N_2} S^{\text{hh}}_{2,i} -   \frac{1}{N_2} \sum_{i=1}^{N_2} S^{\text{hh}}_{1,i},
\end{equation*}
where $S^{\text{snf}}_{1,i} = \sum_{j} y^{\text{mds}}_{Aij}$ and $S^{\text{snf}}_{2,i} = \sum_{j} y^{\text{mds}}_{Bij}$ are the total scores of the observed items in MDS for patients in SNFs on admission and at the later assessment, respectively, $S^{\text{hh}}_{1,i} = \sum_{j} \tilde{y}^{\text{mds}}_{Aij}$ and $S^{\text{hh}}_{2,i} = \sum_{j} \tilde{y}^{\text{mds}}_{Bij}$ are the total scores of the imputed items in MDS for patients receiving home health on admission and at the later assessment, respectively, $N_1$ and $N_2$ are the number of patients treated in SNFs and by home health agencies, respectively, and $N_1 + N_2 = N$. We also define $p_{\text{snf}}$ and $p_{\text{hh}}$ to be the proportion of patients whose functional status improved during the course of the post-acute stay in SNFs and home health care, respectively:
\begin{equation*}
p_{\text{snf}} = \frac{1}{N_1} \sum_{i=1}^{N_1}\bm{1}\{ S^{\text{snf}}_{2,i} < S^{\text{snf}}_{1,i} \}, \quad \text{and} \quad
p_{\text{hh}} = \frac{1}{N_2} \sum_{i=1}^{N_2}\bm{1}\{ S^{\text{hh}}_{2,i} < S^{\text{hh}}_{1,i} \},
\end{equation*}
where $\bm{1}\{ A\}$ is an indicator function that is equal to 1 if $A$ is true and 0 otherwise.

We apply the proposed NMI procedure to examine two quantities: (1) the difference in average change of total scores over the course of post-acute stay between patients in SNFs and those receiving home health; and (2) the difference in proportions of patients whose functional status improved during the post-acute stay between patients in SNFs and those receiving home health, $p_{\text{hh}} - p_{\text{snf}}$. 

Table~\ref{comparison2} displays the point and interval estimates of $d_{\text{hh}} - d_{\text{snf}}$ and $p_{\text{hh}} - p_{\text{snf}}$. The point and interval estimates with nested multiple imputation using either the regression translating model or the MVOPT translating model, as well as with LVM in the Equating step are similar. The results show that on average patients who received home health care do not have a significantly larger functional improvement than those who stayed in SNFs, but more patients who receive home health care improve their functional status during the post-acute stay than those in SNFs. In contrast, the results using the conversion table method suggest that on average patients who received home health care had a significantly higher rate of functional improvement than those who stayed in SNFs. In addition, a larger proportion of patients who received home health care experienced improved functional status in comparison to those who stayed in SNFs.. \citet{gu2016irt} noted that conversion table has a poor performance when it is used to equate MDS and OASIS instruments. Thus, the estimates of the conversion table method in the Equating step may lead to implausible imputations in the Translating step, and overestimation of the rate of functional improvement for patients receiving home health. 


The directions of the point estimates of $d_{\text{hh}} - d_{\text{snf}}$ are different for the different translating step methods, but their interval estimates are partially overlapping. The estimate of $p_{\text{hh}} - p_{\text{snf}}$ for the regression model is larger then that of the MVOPT model, suggesting that different translating models may result in different functional relationship between MDS and OASIS total scores. The MVOPT translating model incorporates more information by relying on all of the items in MDS and OASIS, which should result in a more accurate estimate.


\begin{table}
\centering
\begin{threeparttable}
\caption{Comparison of the estimated differences in average change of total score and estimated differences in proportion of patients whose functional status improve during the course of post-acute stay between patients treated in SNFs to those receiving home health care.}
\label{comparison2}
\begin{tabular}{l ccc c ccc}
\hline
& \multicolumn{3}{c}{$d_{\text{hh}} - d_{\text{snf}}$}  &&  \multicolumn{3}{c}{$p_{\text{hh}} - p_{\text{snf}}$($\%$)} \\
      \cline{2-4} \cline{6-8} 
 & Estimate & SE & 95$\%$ CI  &&  Estimate & SE & 95$\%$ CI \\ 
  \hline
CT\tnote{a}                 &  -1.37 & 0.02  & (-1.42, -1.34)   &&  29.00 & 0.34 & (28.32, 29.66)   \\
LVM\tnote{b}               & -0.01 & 0.05  &  (-0.11, 0.10)     &&  12.08 & 0.53 & (11.02, 13.13) \\
NMI (Reg)\tnote{c} &  0.04  & 0.06  & (-0.08, 0.16)   &&  8.21 & 0.54 & (7.15, 9.26)   \\
NMI (MVOPT)\tnote{d} &  -0.07  & 0.06  & (-0.19, 0.05)   &&  5.93 & 0.55 & (4.84, 7.02)     \\
   \hline
\end{tabular}
\begin{tablenotes}
\item[a] CT: conversion table method;
\item[b] LVM: latent variable matching;
\item[c] NMI (Reg): nested multiple imputation with linear regression translating model;
\item[d] NMI (MVOPT): nested multiple imputation with multivariate ordinal probit translating model.
\end{tablenotes}
\end{threeparttable}
\end{table}

\section{Concluding Remarks}

We proposed a nested multiple imputation procedure to obtain a common patient assessment scale across the continuum of care by imputing unmeasured assessments at multiple dates in two steps. This procedure enables researchers to compare the rates of functional improvement experienced by patients treated in different health care settings using a common measure. This procedure accounts for the uncertainty in both the Equating and Translating steps, and it also provides flexibility for researchers to choose different translating models to impute multiple future assessments without the need to re-equate the instruments. The Equating step utilizes the MVOP model to impute the incomplete instruments that consist of multiple ordinal items. Simulations demonstrated that models based on MVOP are superior to existing methods for imputing incomplete multivariate ordinal variables in most of the experimental conditions that were examined. In addition, including observed covariates improves the point and interval estimates in the Equating step. 

We applied the proposed procedure to analyze patients who had a stroke and were either released home or to SNFs after rehabilitation. Our analyses suggest that more patients who were discharged home and received home health care experience functional improvement in comparison to those who were released to SNFs, but on average the overall functional status improvement across all patients is similar across these settings. This analysis does not imply that one setting is more beneficial to patients than another, because the populations differ in patients' characteristics and initial functional status. However, using the proposed procedure, researchers can identify a subgroup of patients with similar characteristics and initial functional status who were discharged to either of the health care settings, and compare the rates of functional change in this subgroup of patients with the aim of identifying a setting that is more beneficial to certain patients. The proposed procedure can be further extended to impute unmeasured assessments at all assessment dates during patients' post-acute stays. 

The newly proposed methods rely on the conditional independence and the missing at random assumptions. These assumptions are implicitly made in many educational testing applications with the common-items design. Here, these assumptions are somewhat defensible because all three instruments intend to determine the same underlying functional status, and they are all recorded within a close time period. In addition, the proposed methods performed well in a limited simulation analysis in which the two assumptions were violated. Nonetheless, developing procedures that accommodate departures from these assumptions is an area for future research.

One computational limitation of the MVOP model is the complexity of sampling from a truncated multivariate normal distribution, and this complexity is exacerbated when the dimension of the ordinal outcome variables is large. Another computational limitation is sampling the correlation matrix $\mathbf{R}$. Here, we applied a parameter expansion technique to sample $\mathbf{R}$ efficiently. Recent work on using the prior distribution proposed in \citet{lewandowski2009generating} and implemented in Stan \citep{carpenter2016stan} is another possible solution. 


In conclusion, we have proposed a procedure to obtain a common patient assessment scale across the continuum of care. This procedure is flexible and allows researchers to examine the rate of functional improvement using a single instrument.


\begin{supplement}
\stitle{Development of a Common Patient Assessment Scale across the Continuum of Care: A Nested Multiple Imputation Approach}
\slink[url]{http://www.e-publications.org/ims/support/dowload/imsart-ims.zip}
\sdescription{The supplement includes the Slice Sampler Algorithm for the MVOP model, additional results in the simulation study of Section 4, results of posterior predictive checks in Section 5.3 and computer code for an example to illustrate the proposed procedure.}
\end{supplement}


\bibliographystyle{imsart-nameyear}
\bibliography{AOAS_manuscript}






\end{document}